\documentstyle[12pt]{ioplppt}
\begin{document}

\jl{1}

\title{On fusion algebra of chiral $SU(N)_{K}$ models}
\author{A Lima-Santos}

\address{Departamento de Fisica, Universidade Federal de
 S\~ao Carlos, Caixa Postal 676, 13569-905 S\~ao Carlos, Brazil}

\begin{abstract}
We discuss some algebraic setting of chiral $SU(N)_{K}$ models in terms of
the satistical dimensions of their fields. In particular, the conformal
dimensions and the central charge of the chiral $SU(N)_{K}$ models are
calculated from their braid matrices. Furthermore, at level $K=2$, we
present the characteristic polynomials of their fusion matrices in a
factored form.
\end{abstract}

\section{Introduction}

Fusion algebras are found to play an important role in the study of rational
conformal field theories (RCFT). Beside the fact that the fusion rules can
be expressed in terms of the unitary matrix $S$ \cite{Verlinde} that encodes
the modular transformations of the characters of the RCFT 
\begin{equation}
N_{ij}^{k}=\sum_{l}\frac{S_{il}}{S_{0l}}\ S_{jl}\ S_{kl}^{*}  \label{eq1.1}
\end{equation}
where ''$0$ '' refers to the identity operator, and the labels $i,...,l$ run
over n values corresponding to the primary fields of the chiral algebra of
the RCFT, there is a more fundamental reason to look for representations of
the fusion algebra, based on the concept of operator products\cite{BPZ}.
When one tries to compute the operator product coefficients, one is almost
inevitably led to the concept of fusion rules, i.e. formal products 
\begin{equation}
A_{i}\ A_{j}=\sum_{k}N_{ij}^{k}\ A_{k}.  \label{eq1.2}
\end{equation}
of primary fields describing the basis-independent content of the operator
product algebra.

The matrices $N_i$ defined by $(N_i)_{jk}=N_{ij}^k$ form themselves a
representation of the fusion algebra 
\begin{equation}
N_i\ N_j=\sum_kN_{ij}^k\ N_k  \label{eq1.3}
\end{equation}
as follows from unitarity of the matrix $S$; this expresses the
associativity property of the algebra (\ref{eq1.2}). The relation (\ref
{eq1.1}) implies that the matrix $S$ diagonalizes the matrices $N_i$ and
their eigenvalues are of the form 
\begin{equation}
\gamma _i^{(l)}=\frac{S_{il}}{S_{0l}}  \label{eq1.4}
\end{equation}
and obey the sum rules 
\begin{equation}
\gamma _i^{(l)}\ \gamma _j^{(l)}=\sum_kN_{ij}^k\ \gamma _k^{(l)}
\label{eq1.5}
\end{equation}

The general study of these fusion algebras \cite{Kawai} and their
classification have been the object of much work \cite{CPR},\cite{Eholzer},%
\cite{Gepner}.

The numbers 
\begin{equation}
d_{i}\doteq \gamma _{i}^{(0)}=\frac{S_{i0}}{S_{00}}  \label{eq1.6}
\end{equation}
appear as statistical dimensions of superselection sectors \cite{R1} in
algebraic quantum field theory; as square roots of indices for inclusions of
von Neumann algebras \cite{FRS}; as relative sizes of highest weight modules
of chiral symmetry algebras in conformal field theory \cite{Verlinde}; and
in connection with truncated tensor products of quantum groups (see \cite
{Fuchs} for an accomplished review). According to (\ref{eq1.5}), these
numbers obey the statistical dimension sum rules 
\begin{equation}
d_{i}\ d_{j}=\sum_{k}N_{ij}^{k}\ d_{k}.  \label{eq1.7}
\end{equation}
which permits us identify $d_{i}$ as a Frobenius eigenvalue of $N_{i}$.

The paper is organized as follows. In the following section we present the
fusion matrices and the braid matrices for the chiral $SU(N)_{K}$ models.The
conformal dimensions and central charge are calculated from the braid
matrices. In the last section we consider the case $K=2$ and the
characteristic polynomials of the fusion matrices are calculated.

\section{The chiral $SU(N)_{K}$ models}

\subsection{The fusion matrices}

An important fact is that conformal families can be interpreted as
irreducible representations of the chiral algebra. The primary fields of the
chiral $SU(N)_{K}$ models can be represented by those Young tableaux for
which $\stackrel{\sim }{\lambda }_{1}-\stackrel{\sim }{\lambda }_{N}\leq
K,\quad K\geq 2$, where $\stackrel{\sim }{\lambda }_{i}$ is the number of
boxes in the $i$ row. Thus, for a given value of $N$ , the corresponding
Young diagram is a Weyl chamber truncated by the value of $K$.

Let us now associate to each Young tableaux an $(N-1)$-dimensional vector $%
\Lambda $, defined by $\Lambda =\sum_{i=1}^{N-1}\lambda _{i}\ e_{i}$, where $%
\lambda _{i}=\stackrel{\sim }{\lambda }_{i}+N-i$ and $e_{i}$, $i=1,2,...,N$
are the weights in the $N$-dimensional representation of $SU(N)$, satisfying 
$\sum_{i=1}^{N}e_{i}=0$ and $e_{i}\cdot e_{j}=\delta _{ij}-1/N$. The
components of vector $\Lambda $ satisfy $\lambda _{i}\in {\cal N},\quad
\lambda _{1}>\lambda _{2}>\cdots >\lambda _{N-1}>0$ and $N-1\leq \lambda
_{1}\leq N+K-1$.

Proceeding in the $e_{i}$ direction, the next Young tableaux must be
obtained from the preceding one by adding a box in the $i$-row. This gives
us a composition law (fusion rules) for the vectors (primary fields) 
\begin{equation}
\Lambda \pm e_{i}=\left\{ 
\begin{array}{c}
(\lambda _{1},\lambda _{2},\cdots ,\lambda _{i}\pm 1,\cdots ,\lambda
_{N-1})\quad \text{for }i\neq N \\ 
(\lambda _{1}\mp 1,\lambda _{2}\mp 1,\cdots ,\lambda _{N-1}\mp 1)\quad \text{%
for }i=N
\end{array}
\right.  \label{eq1.8}
\end{equation}
i.e., we have in mind an elementary field $\sigma $ (and its conjugate $%
\stackrel{\_}{\sigma }$ ) which interpolates according to the fusion rules 
\cite{Terry} 
\begin{equation}
\lbrack \sigma ][\Lambda ]=\sum_{i=1}^{N}(N_{\sigma })_{\Lambda }^{\Lambda
+e_{i}}[\Lambda +e_{i}],\qquad [\stackrel{\_}{\sigma }][\Lambda
]=\sum_{i=1}^{N}(N_{\stackrel{\_}{\sigma }})_{\Lambda }^{\Lambda
-e_{i}}[\Lambda -e_{i}]  \label{eq1.9}
\end{equation}
The vector $\phi =(N-1,N-2,...,2,1)$ (Weyl vector), may be identified with
the vacuum sector, which contains a vacuum state. In this case the vector $%
\sigma =(N,N-2,\ldots ,2,1)$, is identified with the elementary field and $%
\stackrel{\_}{\sigma }=(N-2,N-3,\ldots ,2,1)$ with its conjugate.

The fusion rules (\ref{eq1.9}) give a natural basis for write the fusion
matrix of elementary fields as: 
\begin{equation}
(N_{\sigma })_{\Lambda }^{\Gamma }=\sum_{i=1}^{N}\widehat{\delta }_{\Lambda
,\Gamma -e_{i}},\qquad (N_{\stackrel{\_}{\sigma }})_{\Lambda }^{\Gamma
}=\sum_{i=1}^{N}\widehat{\delta }_{\Lambda ,\Gamma +e_{i}}  \label{eq1.10}
\end{equation}
where 
\begin{equation}
\widehat{\delta }_{\Lambda ,\Gamma \mp e_{i}}\doteq \left\{ 
\begin{array}{c}
\delta _{\lambda _{1}\gamma _{1}}\delta _{\lambda _{2}\gamma _{2}}\cdots
\delta _{\lambda _{i}\gamma _{i}\mp 1}\cdots \delta _{\lambda _{N-1}\gamma
_{N-1}}\text{, if }i\neq N \\ 
\\ 
\delta _{\lambda _{1}\gamma _{1}\pm 1}\delta _{\lambda _{2}\gamma _{2}\pm
1}\cdots _{\lambda _{N-1}\gamma _{N-1}\pm 1}\text{, if }i=N
\end{array}
\right.  \label{eq1.10a}
\end{equation}

For example, the chiral $SU(2)_{2}$ (Ising) model has three fields, labelled
by $\Phi _{\Lambda }:\Phi _{(1)}=\phi $ (vacuum), $\Phi _{(2)}=\sigma $
(spin) and $\Phi _{(3)}=\epsilon $ (energy) . From the composition law (\ref
{eq1.8}) we can write the fusion rules for the elementary field $\sigma $%
\[
\lbrack \sigma ][\phi ]=[\sigma ],\quad [\sigma ][\sigma ]=[\phi ]+[\sigma
],\quad [\sigma ][\epsilon ]=[\sigma ] 
\]
which give the fusion matrix $\left( N_{\sigma }\right) _{\lambda }^{\gamma
}=\delta _{\lambda \gamma -1}+\delta _{\lambda \gamma +1}$ , and the
remaining fusion matrices can be easily derived by the fusion algebra\ (\ref
{eq1.3}): 
\[
N_{\phi }={\bf 1},\quad N_{\sigma }=\left( 
\begin{array}{ccc}
0 & 1 & 0 \\ 
1 & 0 & 1 \\ 
0 & 1 & 0
\end{array}
\right) ,\quad N_{\epsilon }=\left( 
\begin{array}{ccc}
0 & 0 & 1 \\ 
0 & 1 & 0 \\ 
1 & 0 & 0
\end{array}
\right) . 
\]
For chiral $SU(3)_{K}$ models we have: 
\[
(N_{\sigma })_{\Lambda }^{\Gamma }\equiv (N_{\sigma })_{(\lambda
_{1},\lambda _{2})}^{(\gamma _{1},\gamma _{2})}=\delta _{\lambda _{1}\gamma
_{1}-1}\delta _{\lambda _{2}\gamma _{2}}+\delta _{\lambda _{1}\gamma
_{1}}\delta _{\lambda _{2}\gamma _{2}-1}+\delta _{\lambda _{1}\gamma
_{1}+1}\delta _{\lambda _{2}\gamma _{2}+1}. 
\]
In the case $K=2$ attend six fields $\Phi _{\Lambda }:\Phi _{(2,1)}=\phi $
(vacuum), $\Phi _{(3,1)}=\sigma $, $\Phi _{(4,1)}=\stackrel{\_}{\psi }$, $%
\Phi _{(4,2)}=\epsilon $, $\Phi _{(4,3)}=\psi $ and $\Phi _{(3,2)}=\stackrel{%
\_}{\sigma }$. The fusion rules for the elementary field is 
\begin{eqnarray*}
\lbrack \sigma ][\phi ] &=&[\sigma ],\quad [\sigma ][\stackrel{\_}{\psi }]=[%
\stackrel{\_}{\sigma }],\quad \text{ }[\sigma ][\psi ]=[\epsilon ] \\
\lbrack \sigma ][\stackrel{\_}{\sigma }] &=&[\phi ]+[\varepsilon ],\quad
[\sigma ][\sigma ]=[\psi ]+[\stackrel{\_}{\sigma }],\quad [\sigma ][\epsilon
]=[\stackrel{\_}{\psi }]+[\sigma ].
\end{eqnarray*}
and the fusion matrices are 
\begin{eqnarray}
{\bf \quad N}_{\sigma } &=&\left( 
\begin{array}{cccccc}
0 & 1 & 0 & 0 & 0 & 0 \\ 
0 & 0 & 1 & 1 & 0 & 0 \\ 
1 & 0 & 0 & 0 & 1 & 0 \\ 
0 & 0 & 0 & 0 & 1 & 0 \\ 
0 & 1 & 0 & 0 & 0 & 1 \\ 
0 & 0 & 1 & 0 & 0 & 0
\end{array}
\right) ,N_{\psi }=\left( 
\begin{array}{cccccc}
0 & 0 & 0 & 1 & 0 & 0 \\ 
0 & 0 & 0 & 0 & 1 & 0 \\ 
0 & 1 & 0 & 0 & 0 & 0 \\ 
0 & 0 & 0 & 0 & 0 & 1 \\ 
0 & 0 & 1 & 0 & 0 & 0 \\ 
1 & 0 & 0 & 0 & 0 & 0
\end{array}
\right) ,  \nonumber \\
{\bf N}_{\epsilon } &=&\left( 
\begin{array}{cccccc}
0 & 0 & 0 & 0 & 1 & 0 \\ 
0 & 1 & 0 & 0 & 0 & 1 \\ 
0 & 0 & 1 & 1 & 0 & 0 \\ 
0 & 0 & 1 & 0 & 0 & 0 \\ 
1 & 0 & 0 & 0 & 1 & 0 \\ 
0 & 1 & 0 & 0 & 0 & 0
\end{array}
\right) ,N_{\phi }={\bf 1,\ }N_{\stackrel{\_}{\sigma }}=N_{\sigma }^{t},\ N_{%
\stackrel{-}{\psi }}=N_{\psi }^{t},  \label{eq1.10b}
\end{eqnarray}
where $N^{t}$ standing for the transposed of $N$.

Inspecting the Young diagrams for a given $N$ and a given value of $K$, one
can solve (\ref{eq1.7}) to get the statistical dimensions associated with
each irreducible representation of the $SU(N)_{K}$ models\cite{Kac},\cite
{Fuchs1}:

\begin{equation}
d(\Lambda )=\prod_{i=1}^{N-1}\frac{s(\lambda _{i})}{s(1)}\prod_{i<j}^{N-1}%
\frac{s(\lambda _{i}-\lambda _{j})}{s(j-i+1)}  \label{eq1.11}
\end{equation}
where $s(\lambda )=\sin (\pi \lambda /(N+K))$. In particular, $d(\phi )=1$
and $d(\sigma )=s(N)/s(1).$ The statistical dimensions of conjugate
representations coincide.

\subsection{The exchange algebra}

Due to the non-additivity of conformal scale dimensions, the spectrum
decomposition of local fields $\Phi (\stackrel{\rightarrow }{x})$, with
respect to the center of the conformal group is non-trivial\cite{RS2}: 
\begin{equation}
\Phi (\stackrel{\rightarrow }{x})=\sum_{\eta }\Phi _{\eta }(\stackrel{%
\rightarrow }{x}),  \label{eq1.12}
\end{equation}
where every $\Phi _{\eta }(\stackrel{\rightarrow }{x})$ is a non-local
object with $\eta $-dependent complex phases that occur in the special
conformal transformation laws. The range of the label $\eta $ is determined
by the selection rules of scale dimensions.

Applied to the vacuum state, the fields of a conformal block $[\alpha ]$
generate a representation sector ${\cal H}_{\alpha }$ of the stress-energy
tensor field. Applied to a state in ${\cal H}_{\beta }$, fields of a
conformal block $[\alpha ]$ give us contributions in all space ${\cal H}%
_{\gamma }$ allowed by the fusion rules. Introducing orthogonal projectors $%
P_{\beta }$ on the sectors ${\cal H}_{\beta }$ one obtains the decomposition 
\begin{equation}
\Phi ^{\alpha }(\stackrel{\rightarrow }{x})=\sum_{\beta ,\gamma }P_{\gamma
}\Phi ^{\alpha }(\stackrel{\rightarrow }{x})P_{\beta }\equiv (\Phi ^{\alpha
})_{\gamma \beta }(\stackrel{\rightarrow }{x}).  \label{eq1.13}
\end{equation}
This decomposition coincides with the spectral decomposition (\ref{eq1.12})
with the previous label $\eta $ replaced by '' fusion channels '' for the
''charge '' $\alpha $.

It is a well established fact that conformal field theories can be
constructed on Hilbert spaces which are direct sums of irreducible
representations of an observable algebra ${\cal L}\oplus \stackrel{\_}{\cal L%
}$. Both subalgebras ${\cal L}\oplus {\bf 1}$ and ${\bf 1\oplus }\stackrel{\_%
}{\cal L}$ is associated to one light-cone. We also add the further
requirement that the Hilbert space contains only a finite number of
irreducible representations of ${\cal L}$ and $\stackrel{\_}{\cal L}$. Hence 
\begin{equation}
{\cal H}=\oplus _{\alpha ,\stackrel{\_}{\alpha }}{\cal H}_{\alpha }\otimes 
{\cal H}_{\stackrel{\_}{\alpha }}  \label{eq1.14}
\end{equation}
where ${\cal H}_{\alpha }$ (${\cal H}_{\stackrel{\_}{\alpha }}$) are
irreducible representations of ${\cal L}$ ($\stackrel{\_}{\cal L}$) and the
pair ($\alpha ,\stackrel{\_}{\alpha }$) takes its values in a finite set.

Due to the light-cone factorization of the stress-energy tensor field
algebra, the label $[\alpha ]$ of conformal blocks are in fact pairs $%
[\alpha _{+},\alpha _{-}]$. Both representation sectors and the projectors
factorize into the projected fields 
\begin{equation}
(\Phi ^{\alpha })_{\gamma \beta }(\stackrel{\rightarrow }{x})=(A^{\alpha
_{+}})_{\gamma _{+}\beta _{+}}(\stackrel{\rightarrow }{x}_{+})\otimes
(A^{\alpha _{-}})_{\gamma _{-}\beta _{-}}(\stackrel{\rightarrow }{x}_{-}).
\label{eq1.15}
\end{equation}
Finally the monodromy properties of the conformal blocks are equivalent to
the exchange algebra on either light-cone 
\begin{equation}
(A^{\alpha _{1}})_{\delta \gamma }(x)(A^{\alpha _{2}})_{\gamma \beta
}(y)=\sum_{\gamma ^{\prime }}\left[ R_{(\alpha _{1},\alpha _{2})}^{(\delta
,\beta )}(s)\right] _{\gamma \gamma ^{\prime }}(A^{\alpha _{2}})_{\delta
\gamma ^{\prime }}(y)(A^{\alpha _{1}})_{\gamma ^{\prime }\beta }(x).
\label{eq1.16}
\end{equation}
Here (and from now on ) we have omitted the indices $"\pm ".$ The numerical
structure constants $R$ are matrices which satisfy three basic properties
(see \cite{RS1}):

(i) $\left[ R_{(\alpha _{1},\alpha _{2})}^{(\delta ,\beta )}(s)\right] $
depend on $x$ and $y$ only through their relative position. This follows
from translation and scale variance. Moreover, if $s=sign(x-y)=\pm $, then 
\begin{equation}
\left[ R_{(\alpha _{1},\alpha _{2})}^{(\delta ,\beta )}(+)\right]
^{-1}=\left[ R_{(\alpha _{1},\alpha _{2})}^{(\delta ,\beta )}(-)\right]
\label{eq1.17}
\end{equation}

(ii) {\em Phase condition: }$R_{(\alpha _{1},\alpha _{2})}^{(\delta ,\beta
)} $ and $R_{(\alpha _{2},\alpha _{1})}^{(\delta ,\beta )}$ are related
through the following relation 
\begin{equation}
\sum_{\gamma ^{\prime }}[R_{(\alpha _{1},\alpha _{2})}^{(\delta ,\beta
)}(s)]_{\gamma \gamma ^{\prime }}[R_{(\alpha _{2},\alpha _{1})}^{(\delta
,\beta )}(s)]_{\gamma ^{\prime }\gamma ^{\prime \prime }}\exp (2i\pi
(h_{\gamma }+h_{\gamma ^{\prime }}-h_{\delta }-h_{\beta })=\delta _{\gamma
,\gamma ^{\prime \prime }}.  \label{eq1.18}
\end{equation}
Where $h_{\gamma }$ 's are primary dimensions of the representations $%
[\gamma ]$. This follows from invariance under special conformal
transformation.

(iii) {\em Braid relations: }The exchange matrices satisfy 
\[
\sum_{\beta _{1}^{\prime \prime }}[R_{(\alpha _{1},\alpha _{2})}^{(\beta
_{0},\beta _{2})}(s)]_{\beta _{1}\beta _{1}^{\prime \prime }}\ [R_{(\alpha
_{1},\alpha _{3})}^{(\beta _{1}^{\prime \prime },\beta _{3})}(s)]_{\beta
_{2}\beta _{2}^{\prime \prime }}\ [R_{(\alpha _{2},\alpha _{3})}^{(\beta
_{0},\beta _{2}^{\prime })}(s)]_{\beta _{1}^{\prime \prime }\beta
_{1}^{\prime }}= 
\]
\begin{equation}
\sum_{\beta _{2}^{\prime \prime }}[R_{(\alpha _{2},\alpha _{3})}^{(\beta
_{1},\beta _{2})}(s)]_{\beta _{2}\beta _{2}^{\prime \prime }}\ [R_{(\alpha
_{1},\alpha _{3})}^{(\beta _{0},\beta _{2}^{\prime \prime })}(s)]_{\beta
_{1}\beta _{1}^{\prime }}\ [R_{(\alpha _{1},\alpha _{2})}^{(\beta
_{1}^{\prime },\beta _{3})}(s)]_{\beta _{2}^{\prime \prime }\beta
_{2}^{\prime }},  \label{eq1.19}
\end{equation}
which is the consistency relation for the associativity of the exchange
algebra (\ref{eq1.16}).

All these relations were derived in \cite{FRS} from the theory of localized
endomorphism without invoking conformal invariance.

The solution of eq.(\ref{eq1.19}) for all admissible (i.e. consistent with
the fusion rules (\ref{eq1.9}) ) indices being irreducible representations
of $SU(N)_{K}$ and \ $\alpha _{1}=\alpha _{2}=\alpha _{3}=\sigma $,
''elementary field '' \cite{Pasquier} , is given by 
\begin{eqnarray}
\left[ R_{(\sigma ,\sigma )}^{(\Lambda +e_{k}+e_{s},\Lambda )}(+)\right]
_{\Lambda +e_{k},\Lambda +e_{s}} &=&\eta \ q^{\frac{1}{2}}\left\{ \frac{%
s(\lambda _{k}-\lambda _{s}+1)s(\lambda _{k}-\lambda _{s}-1)}{s(\lambda
_{k}-\lambda _{s})^{2}}\right\} ^{1/2}  \nonumber \\
\left[ R_{(\sigma ,\sigma )}^{(\Lambda +e_{k}+e_{s},\Lambda )}(+)\right]
_{\Lambda +e_{k},\Lambda +e_{k}} &=&\eta \ q^{\frac{1-\lambda _{k}+\lambda
_{s}}{2}}\left\{ \frac{s(1)}{s(\lambda _{k}-\lambda _{s})}\right\} \text{, 
{\rm for } }k\neq s  \nonumber \\
\left[ R_{(\sigma ,\sigma )}^{(\Lambda +2e_{k},\Lambda )}(+)\right]
_{\Lambda +e_{k},\Lambda +e_{k}} &=&-\ \eta \ q  \label{eq1.20}
\end{eqnarray}
where $\Lambda =(\lambda _{1},\lambda _{2},...,\lambda _{N-1})$ , $s(\lambda
)=\sin (\pi \lambda /(N+K))$ , $q=\exp (-\frac{2i\pi }{N+K})$ and $\eta $ is
a arbitrary phase factor which will be fixed latter.

\subsection{The dimensional trajectories and central charge}

The $R$ matrices (\ref{eq1.20}) have been constructed as solution of the
braid relations (\ref{eq1.19}) for the elementary field $\sigma $, of the
chiral $SU(N)_{K}$ models. They must also solve the phase condition (\ref
{eq1.18}). This yields constraints 
\begin{equation}
\eta ^{-2N}=q^{N+1}  \label{eq1.22}
\end{equation}
as well as the following equations for the dimensional trajectories 
\begin{eqnarray}
\exp \{2i\pi (2h_{\Lambda +e_{k}}-h_{\Lambda }-h_{\Lambda +2e_{k}})\}
&=&\eta ^{-2}q^{-2},  \nonumber \\
\exp \{2i\pi (2h_{\Lambda +e_{k}}-h_{\Lambda }-h_{\Lambda +e_{k}+e_{s}})\}
&=&\eta ^{-2},  \label{eq1.23}
\end{eqnarray}
for $\lambda _{k}-\lambda _{s}=1$. We have two furthermore equations when $%
\mid \lambda _{k}-\lambda _{s}\mid =2$, which are 
\begin{eqnarray}
\exp \{2i\pi (2h_{\Lambda +e_{k}}-2h_{\Lambda +e_{s}})\} &=&q^{-2(\lambda
_{k}-\lambda _{s})},  \nonumber \\
\exp \{2i\pi (2h_{\Lambda +e_{k}}-h_{\Lambda }-h_{\Lambda +e_{k}+e_{s}})\}
&=&\eta ^{-2}q^{-(\lambda _{k}-\lambda _{s})-1}.  \label{eq1.24}
\end{eqnarray}

Eq.(\ref{eq1.22}) together with the normalized $2$-point function of the
elementary field allows us to choose among all possible solutions, one that
fixes the value of the phase factor, $\eta $, to be related with the
conformal dimension of the elementary field $\sigma $: 
\begin{equation}
\eta =\exp \{-\frac{2i\pi }{N-1}h_{\sigma }\}  \label{eq1.25}
\end{equation}

Next, using the Kac-determinante\cite{Kac1} , we assume that the conformal
spectrum of the chiral $SU(N)_{K}$ models can be derived from (\ref{eq1.23}
and \ref{eq1.24}) by the ansatz 
\begin{equation}
h_{\Lambda }=\sum_{k=1}^{N-1}(a_{k}\lambda _{k}^{2}+b_{k}\lambda
_{k})+\sum_{k<s}^{N-1}c_{ks}\lambda _{k}\lambda _{s}+d  \label{eq1.26}
\end{equation}
where $a_{k},b_{k},c_{ks}$ and $d$ are functions of $N$ and $K$.

Therefore the primary dimensions ($\bmod{Z}$) of the irreducible
representations of the chiral $SU(N)_{K}$ models\cite{Pasquier} can be
written as: 
\begin{eqnarray}
h_{\Lambda } &=&\frac{(N+K-1)(N-1)}{2N(N+K)}\sum_{i=1}^{N-1}\lambda _{i}^{2}-%
\frac{(N+K-1)}{N(N+K)}\sum_{i<j}^{N-1}\lambda _{i}\lambda _{j}  \nonumber \\
&&\ +\sum_{i=1}^{N-1}(i-\frac{N+1}{2})\lambda _{i}+\frac{(N+K+1)N(N^{2}-1)}{%
24(N+K)}  \label{eq1.27}
\end{eqnarray}

It is also easy to verify that the $h_{\Lambda }$-dimensions enjoy the
symmetry 
\begin{equation}
h_{(\lambda _{1},\lambda _{2},\cdots ,\lambda _{N-1})}=h_{(\lambda
_{1},\lambda _{1}-\lambda _{N-1},\lambda _{1}-\lambda _{N-2},\cdots ,\lambda
_{1}-\lambda _{2})}
\end{equation}
which reflects the $Z_{N}$-symmetry generated by $e_{i}\rightarrow e_{i+1}$, 
$i=1,2,...,N-1$ and $e_{N}\rightarrow e_{1}$. It means that the conformal
dimension of conjugate representations coincide.

According to references \cite{FRS,R1} we introduce the statistics phase $%
\omega (\Lambda )$, which generalizes the distinction between bosons and
fermions of para-statistics, and putting $\alpha =\sum d^{2}(\Lambda )\omega
^{-1}(\Lambda )$ one can define the matrices $S$ and $T$ by:

\begin{eqnarray}
S_{\Lambda \Gamma } &=&\frac{1}{\left| \alpha \right| }\sum_{\Lambda
^{\prime }}N_{\Lambda \Gamma }^{\Lambda ^{\prime }}\frac{\omega (\Lambda
)\omega (\Gamma )}{\omega (\Lambda ^{\prime })}d(\Lambda ^{\prime }),
\label{eq1.28a} \\
T &=&\left( \frac{\alpha }{\left| \alpha \right| }\right) ^{1/3}\text{{\rm %
Diag}}(\omega (\Lambda )),  \label{eq1.28b}
\end{eqnarray}
which satisfy 
\begin{eqnarray}
SS^{\dagger } &=&TT^{\dagger }={\bf 1}_{N}\ ,\quad TSTST=S,  \nonumber \\
S^{2} &=&C,\quad \ CT=TC=T,  \label{eq1.29}
\end{eqnarray}
where $C_{\Lambda \Gamma }=\delta _{\stackrel{\_}{\Lambda }\Gamma }$ is the
conjugation matrix. This algebra is famous from RCFT but, as observed in 
\cite{FRS}, it does not depend on any covariance or modular properties.

For the chiral $SU(N)_{K}$ models, the statistics phase is defined by $%
\omega (\Lambda )=\exp (2i\pi h_{\Lambda })$ (Spin-Statistics Theorem\cite
{FRS}), where $h_{\Lambda }$ is the conformal dimension of the primary field 
$\Lambda $ given by (\ref{eq1.27})

Invariance under $SL(2,C)$ transformations allows one to derive from eq.(\ref
{eq1.28b}) an interesting relation between the central charge and the
statistical dimensions \cite{R1}: 
\begin{equation}
\exp (2i\pi \frac{c}{8})=\frac{\sum_{\Lambda }d^{2}(\Lambda )\exp (2i\pi
h_{\Lambda })}{\sqrt{\sum_{\Lambda }d^{2}(\Lambda )}}.  \label{eq1.30}
\end{equation}

Substituting (\ref{eq1.11}) and (\ref{eq1.27}) into (\ref{eq1.28b}) we get
the central charge ($\bmod{Z}$) for $SU(N)_{K}$ models 
\begin{equation}
c=(N-1)\left( 1-\frac{N(N+1)}{(N+K)(N+K-1)}\right) .  \label{eq1.31}
\end{equation}

\section{The $SU(N)_{2}$ models}

At the level $K=2$ this central charge (\ref{eq1.31}) reduces to $%
c=2(N-1)/(N+2)$ and the primary fields $\Lambda $ are identified with the
order fields $\sigma _{k}$ , $k=1,2,...,N-1$, $Z_{N}$-neutral fields $%
\epsilon ^{(j)},j=1,2,...\leq N/2$ and the parafermionic currents $\Psi _{k}$%
, $k=0,1,...,N-1$ , in Zamolodchikov-Fateev's parafermionic theories\cite
{Zamo1}. For each $\Lambda $-field we define a ''charge ''{\em \ } $\nu
=\sum_{i=1}^{N-1}\lambda _{i}-N(N-1)/2\quad \bmod{2N}$, and collect these $%
N(N+1)/2$ primary fields in $N$ {\em cominimal equivalence classes }\cite
{NRS}, $[\phi _{k}^{k}]$ , $k=0,1,...,N-1,$ according to their statistical
dimensions (\ref{eq1.11}): 
\begin{eqnarray}
d_{k} &=&\prod_{i=0}^{k-1}\frac{s(N-i)}{s(i+1)},\qquad s(x)=\sin (\frac{x\pi 
}{N+2})  \nonumber \\
d_{0} &=&1,\qquad d_{N-k}=d_{k},\qquad k=1,2,...,N-1  \label{eq1.32}
\end{eqnarray}

$SU(N)_{2}$ representations of the order fields $\phi _{k}^{k}$ , $%
k=1,...,N-1$ are the fully antisymmetric Young tableaux with $k$ boxes.
Tableaux of fields comprising a cominimal equivalence class $\phi _{\nu
}^{k} $ in which the representation $\phi _{k}^{k}$ appears, $(\nu =k\ \bmod{%
2}$ , i.e., $\nu =k,k+2,\cdots ,2N-2-k)$, are obtained by adding $(\nu -k)/2$
rows of width $2$ to the top of the reduced tableau of $\phi _{k}^{k}$ .

These equivalence classes are generated by $Z_{N}$ symmetry which connect
the representations belonging to each class through of the fusion rules\cite
{Gepner1} 
\begin{equation}
\phi _{\nu _{1}}^{k_{1}}\times \phi _{\nu
_{2}}^{k_{2}}=\sum_{k=|k_{1}-k_{2}|\bmod{2}}^{{\min }%
(k_{1}+k_{2},2N-k_{1}-k_{2})}\phi _{\nu _{1}+\nu _{2}}^{k}  \label{eq1.33}
\end{equation}
In particular, the elementary field $\phi _{1}^{1}$, $(\phi _{1}^{1}\times
\phi _{\nu }^{k}=\phi _{\nu +1}^{k-1}+\phi _{\nu +1}^{k+1})$ connects the
equivalence class of $\phi _{\nu }^{k}$ with adjacent classes, while the
field $\phi _{2}^{0}$ , $(\phi _{2}^{0}\times \phi _{\nu }^{k}=\phi _{\nu
+2}^{k})$, connects the fields in the same cominimal equivalence class.
Thus, the chiral $SU(N)_{2}$ fusion algebra can be generated by these two
fields. For example, the $6$ primary fields of $SU(3)_{2}$ can be collected
in $3$ cominimal equivalence classes ( modulo the identification $\phi _{\nu
}^{k}=\phi _{N+\nu }^{N-k}$ ) as: 
\begin{equation}
\left\{ 
\begin{array}{lllllll}
&  & \phi _{2}^{2} &  &  & \rightarrow & d_{2}=\frac{s(2)}{s(1)} \\ 
&  &  &  &  &  &  \\ 
& \phi _{1}^{1} &  & \phi _{3}^{1} &  & \rightarrow & d_{1}=\frac{s(3)}{s(1)}
\\ 
&  &  &  &  &  &  \\ 
\phi _{0}^{0} &  & \phi _{2}^{0} &  & \phi _{4}^{0} & \rightarrow & d_{0}=%
\frac{s(4)}{s(1)}
\end{array}
\right\} .  \label{eq1.34}
\end{equation}

\subsection{The characteristic polynomials}

The fusion rule for the field $\phi _{2}^{0}$ with any primary field $\phi
_{\nu }^{k}$ has only one term on the right-hand side. Such fields indicate
that the fusion rules can be naturally represented as the chiral ring of
some perturbed topological Landau-Ginzburg theory \cite{Intriligator}, and
the correspondent ''potential '' is obtained by integrate some constraint
equations\cite{Gepner2}, $P_{i}(x_{1},x_{2},...,x_{n})=0$. If there exists
at least one fusion matrix non degenerate, $N_{f}$, i.e., with non
degenerate eigenvalues, any fusion matrix may be written as $N_{i}=$ $%
P_{i}(N_{f})$. The matrix $N_{f}$, on the other hand, satisfies its
characteristic equation $P(x)=0$, that is also its minimal equation\cite
{Zuber}.The constraint on $N_{f}$ is thus $P(N_{f})=0$ that can be
integrated to yield a ''potential ''.

Here we proceed to the explicit computation of the characteristic
polynomials $P_{\nu }^{k}(x)=\det (x{\bf 1}-N_{\phi _{\nu }^{k}})$
associated with the primary fields of the $SU(3)_{2}$ model. Using the
fusion matrices $N_{\phi _{\nu }^{k}}$ , given by (\ref{eq1.10b}) we get: 
\begin{eqnarray}
P_{0}^{0}(x) &=&(x-1)^{6}  \nonumber \\
P_{2}^{0}(x) &=&P_{4}^{0}(x)=x^{6}-2x^{3}+1  \nonumber \\
P_{1}^{1}(x) &=&P_{2}^{2}(x)=x^{6}-4x^{3}-1  \nonumber \\
P_{3}^{1}(x) &=&x^{6}-3x^{5}+5x^{3}-3x-1  \label{eq1.35}
\end{eqnarray}

Now we introduce the numbers 
\begin{equation}
d_{k}(n)=\frac{\sin (\frac{n(4-k)\pi }{5})}{\sin (\frac{n\pi }{5})},\quad
k=0,1,2,\quad n=1,2.  \label{eq1.36}
\end{equation}
to write (\ref{eq1.35}) in a factored form

\begin{eqnarray}
P_{2}^{0}(x) &=&P_{4}^{0}(x)=(x^{3}-d_{0}^{3}(1))(x^{3}+d_{0}^{3}(2)) 
\nonumber \\
P_{1}^{1}(x) &=&(x^{3}-d_{1}^{3}(1))(x^{3}-d_{1}^{3}(2))  \nonumber \\
P_{3}^{1}(x) &=&(x-d_{1}(1))^{3}(x-d_{1}(2))^{3}  \nonumber \\
P_{2}^{2}(x) &=&(x^{3}-d_{2}^{3}(1))(x^{3}+d_{2}^{3}(2)).  \label{eq1.37}
\end{eqnarray}

This construction is extended for all $SU(N)_{2}$ models. For each
irreducible representation $\phi _{\nu }^{k}$ we associate a factored
characteristic polynomial which depend on the parafermionic charge $\nu $
according to $\frac{N}{\nu }=\frac{p}{q}$ where $p$ and $q$ are positive
integers mutually coprime: 
\begin{equation}
P_{\nu }^{k}(x)=\prod_{n=1}^{\frac{N+1}{2}}\left( x^{p}-d_{k}^{p}(n)\right)
^{\frac{\nu }{q}},\text{ \quad {\rm if\quad }}p.q\text{{\rm -odd}}
\end{equation}
\begin{equation}
P_{\nu }^{k}(x)=\prod_{n=1}^{\frac{N+1}{2}}\left(
x^{p}+(-1)^{n}d_{k}^{p}(n)\right) ^{\frac{\nu }{q}},\text{ \quad {\rm %
if\quad }}p.q\text{{\rm -even}}
\end{equation}
for $N$-odd, and 
\begin{equation}
P_{\nu }^{k}(x)=\left( x^{p}-d_{k}^{p}(l)\right) ^{\frac{\nu }{2q}%
}\prod_{n=1}^{\frac{N}{2}}\left( x^{p}-d_{k}^{p}(n)\right) ^{\frac{\nu }{q}},%
\text{ \quad {\rm if\quad }}p.q\text{{\rm -odd}}
\end{equation}
\begin{equation}
P_{\nu }^{k}(x)=\left( x^{p}+(-1)^{l}d_{k}^{p}(l)\right) ^{\frac{\nu }{2q}%
}\prod_{n=1}^{\frac{N}{2}}\left( x^{p}+(-1)^{n}d_{k}^{p}(n)\right) ^{\frac{%
\nu }{q}},\text{ \quad {\rm if\quad }}p.q\text{{\rm -even}}
\end{equation}
where $l=(N+2)/2$, for $N$-even.

Here we have introduced the numbers 
\begin{eqnarray}
d_{k}(n) &=&\frac{\sin (\frac{n(N+1-k)\pi }{N+2})}{\sin (\frac{n\pi }{N+2})}%
,\quad k=0,1,2,...,N-1,  \nonumber \\
\quad n &=&1,2,...,\leq \frac{N+2}{2}  \label{eq1,42}
\end{eqnarray}
which satisfy the following sum rules 
\begin{equation}
d_{i}(n)d_{j}(n)=\sum_{k}(N_{i})_{j}^{k}\ d_{k}(n).  \label{eq1.43}
\end{equation}

At level $K>$ $2$ , it is also possible write the characteristic polynomials
of the fusion matrices of the chiral $SU(N)_{K}$ models. In particular, for
the elementary field $\sigma $, the fusion matrix is given by (\ref{eq1.10})
and we can use (\ref{eq1.4}) and (\ref{eq1.28a}) to write the correspondent
characteristic polynomial as: 
\begin{equation}
P_{\sigma }(x)=\prod_{\Lambda }\left( x-\frac{\exp \{2i\pi (h_{\Lambda
}+h_{\sigma })\}}{d(\Lambda )}\sum_{k=1}^{N}\frac{d(\Lambda +e_{k})}{\exp
\{2i\pi (\Lambda +e_{k})\}}\right) ,
\end{equation}
where $h_{\Lambda }$ is the conformal dimension of the field $\Lambda $,
given by (\ref{eq1.27}), and $d(\Lambda )$ its statistical dimension, given
by (\ref{eq1.11}). For the other fields, the expression for their
characteristic polynomials are more complex.

{\bf \flushleft  Acknowledgments. }I would like to thank Prof. Roland
K\"{o}berle for useful discussions. 

\end{document}